\documentclass[draft]{agujournal2019}
\draftfalse
\journalname{Space Weather}
\usepackage[inline]{trackchanges}

\usepackage{graphicx,subcaption}
\usepackage{booktabs,multirow,multicol}
\usepackage{lineno}

\usepackage{amsmath}

\usepackage{stackengine}
\usepackage{adjustbox}

\usepackage{url}
\usepackage{soul}

\newcommand{\tR}{t_{\mathrm{R0}}}
\newcommand{\tA}{t_{\mathrm{A}}}
\newcommand{\tsyn}{t_{\mathrm{syn}}}
\newcommand{\TP}{T_{\mathrm{P}}}
\newcommand{\TF}{T_{\mathrm{F}}}
\newcommand{\TL}{T_{\mathrm{L}}}
\newcommand{\NR}{N_{\mathrm{R}}}
\newcommand{\NA}{N_{\mathrm{A}}}
\newcommand{\MSD}{\mathrm{MSD}}
\newcommand{\NRMSE}{\mathrm{NRMSE}}
\newcommand{\Skill}{\mathrm{Skill}}
\newcommand{\FFT}{\mathrm{FFT}}
\newcommand{\invFFT}{\mathrm{invFFT}}
\newcommand{\SR}{\mathrm{SR}}
\newcommand{\fitSR}{\mathrm{f}_{\mathrm{SR}}}
\newcommand{\slopeSR}{\gamma_{\mathrm{SR}}}
\newcommand{\HFSR}{\delta_{\mathrm{SR}}}
\newcommand{\freq}{f}

\newcommand{\predict}{\Pi}
\newcommand{\corrtime}{\tau_{\mathrm{c}}}
\newcommand{\corrlength}{\lambda_{\mathrm{c}}}

\newcommand{\valueFLT}[1]{#1}
\newcommand{\decade}[1]{10^{#1}}
\newcommand{\valueSI}[2]{\valueFLT{#1} \times \decade{#2}}
\newcommand{\unit}[1]{~ \mathrm{#1}}

\begin{document}

\title{Analog ensemble forecasts of solar wind parameters: Quantification of the predictability and time-domain spectral performance}

\authors{
  Pauline A. Simon\affil{1}, 
  Christopher H. K. Chen \affil{1},
  Mathew J. Owens \affil{2},
  and
  Chaitanya Sishtla \affil{1}
}
\affiliation{1}{Department of Physics and Astronomy, Queen Mary University of London, London, UK}
\affiliation{2}{Department of Meteorology, University of Reading, Reading, UK}
\correspondingauthor{Pauline A. Simon}{pauline.simon@qmul.ac.uk}

\begin{keypoints} 
\item The Analog Ensemble (AnEn) method can forecast mesoscale structures of solar wind velocity and magnetic field for resolutions higher than 1h.
\item AnEn forecasts can be more time-accurate than persistence, climatology, and synodic recurrence for solar wind velocity and magnetic field.
\item An ensemble reduction based on the spectra
of the AnEn forecasts preserves the forecast of the fluctuations in the reduction process.
\end{keypoints}

\subsection*{Key Words:}
space weather, solar wind, analog ensemble, forecasting, WIND, mesoscale fluctuations

\abstract{
    Forecasting multiscale properties of the solar wind is one of the important aspects of space weather prediction as mesoscales, larger than one minute, can affect the magnetosphere. Amongst forecasting techniques, the Analog Ensemble (AnEn) method allows the forecast of a quantity from its past behavior, is easy and quick to implement, and results in an ensemble of time series.
    A comparison of optimal AnEn forecasts of \textit{Wind} spacecraft observations of near-Earth solar wind properties with the persistence and climatology baselines allows a quantification of the predictability of the magnetic and velocity components and magnitude. The AnEn predictions were found to be as accurate as persistence for short-term forecasts and climatology for long-term ones, and performed better than both baselines for more than 60\% of the samples for a particular lead time. Furthermore, using an AnEn instead of the baselines enables prediction of the full spectrum of solar wind fluctuations. However, using the standard averaging method to generate a unique forecast from the AnEn ensemble results in a loss of power in the small-scale fluctuations. To prevent this loss, a new spectral reduction method is proposed and compared to the standard averaging method as well as the synodic recurrence baseline. The AnEn spectral-reduced forecast is shown to be more time-accurate than the synodic baseline and more frequency-accurate than the mean-reduced forecasts. Such a reduced forecast is then confirmed to be useful as a comparative baseline in performance diagnostics of space weather models.}


\subsection*{Plain Language Summary} 
Forecasting solar wind behavior is crucial for space weather applications, as it impacts the Earth's magnetic environment and our society. The role of fluctuations of the velocity and the magnetic field, of duration larger than one minute, is important in this impact. Amongst forecasting techniques, the Analog Ensemble (AnEn) method can forecast a quantity from its past behavior, is easy and quick to implement, and results in an ensemble of time series. Comparing the AnEn forecasts of the solar wind observations at half-minute resolution with alternate forecasts shows the effectiveness of AnEn in forecasting more than half of our dataset and estimating temporal performance. However, smaller-scale fluctuations are lost while making an average to reduce the forecast ensemble to one time series. To preserve such fluctuations, we propose a methodology based on the spectral properties of the ensemble and confirm its performance with a novel spectral diagnostic. The new reduced forecast can be useful to replace missing data in solar wind measurements, to inform solar wind models or to forecast the magnetospheric activity. The post-treatment spectral techniques discussed here are also applicable beyond the field of Space Weather. 


\section{Introduction}
The recent geomagnetic storm of 10 May 2024, triggered by a series of interplanetary coronal mass ejections (ICME) \cite{liu_pileup_2024}, confirmed that current space weather models fall short on multiple fronts \cite{hayakawa_solar_2025}. Inaccurate predictions during the storm, including poor forecasts of geomagnetic indices \cite{parker_satellite_2024}, and the post-application of the currently used models to that event -- for instance magnetospheric \cite{tulasi_ram_super-intense_2024} -- highlight shortcomings in our understanding of the formation and acceleration of solar wind plasma, its evolution in the interplanetary space, its interactions with the magnetosphere, and the magnetic structure of ICMEs. Such improvements are critical given our increasing reliance on technology sensitive to space weather disruptions \cite{eastwood_economic_2017, oughton_risk_2019}.

Space weather forecasting can be viewed as a coupling between different systems. Measurements at the first Lagrangian point (L1), $0.01\unit{AU}$ (astronomical unit) upstream of the Earth, are used by inner-magnetospheric and ionospheric models to forecast geomagnetic activity. Thus, lead times of such forecasts are limited to less than an hour as solar wind plasma at L1 reaches Earth in less than an hour. To increase that lead time, L1 measurements need to be forecast. Coronal model output initialized with remote measurements of the solar activity \cite{owens_metrics_2008}, and in-situ measurements of sub-L1 solar wind parameters \cite{lugaz_need_2025}, propagated to L1 by solar wind models can increase the lead time to several days. However, improving the accuracy of the forecasts requires improvements in each step of the modelling pipeline. \citeA{macneice_assessing_2018} provides an overview of the state-of-the-art modelling efforts to forecast solar wind conditions at $1\unit{AU}$. The currently obtained errors of L1 forecasts come from various sources: the boundary conditions close to the sun, the background solar wind structures and ongoing interactions \cite{hinterreiter_assessing_2019} or the under-resolved grids used by the physics-based models \cite{regnault_discrepancies_2024}. The physics-based models of the solar wind reflects mostly large scales in accordance to an extensible literature associated to the geomagnetic impact of large scales structures \cite{dremukhina_relationship_2018}. However, the multi-scale and non-linear behavior of the solar wind can be observed in the Fourier spectral space and is spread through more than 8 decades of frequencies \cite{roberts_nature_1987,sahraoui_magnetohydrodynamic_2020}, from days and hours (scales of ICME and Corotating Interaction Regions, here referred to large scales), to the scales associated to proton and electrons smaller than $10\unit{s}$ (here referred to small scales). In particular, mesoscale structures larger than a minute have the right size to affect Earth's environment in a quasi-stationary way \cite{viall_mesoscale_2021}. For instance, the periodic density structures are associated to various spatial mesoscales \cite{kepko_inherent_2020} and can be correlated to enhancements of the magnetic power spectra \cite{di_matteo_azimuthal_2024}. Furthermore, as the coupling between the solar wind and magnetosphere is not linear, with energy loading and abrupt release, an accumulation of small perturbations can generate geomagnetic activity \cite{damicis_effect_2020}.  \citeA{owens_ensemble_2014} demonstrate the efficiency of a well-fitted noise added to solar wind forecasts in the improvement of a magnetospheric model's responses.  A recent study \cite{ala-lahti_impact_2024} concludes that  $15\unit{min}$ and longer fluctuations (mesoscale) are needed to well estimate the starting point of a substorm and capture $40\unit{\%}$ of the energy transfer, while higher frequency fluctuations ($2$ to $8\unit{min}$ Ultra-Law-Frequency (ULF) waves) contribute up to $15\unit{\%}$ of the energy transfer. Note that these large and mesoscale ($> 15\unit{min}$) fluctuations are what is defined as ``low frequency" fluctuation by \citeA{sibeck_quantifying_2025}. One way to introduce mesoscale fluctuations to large-scale L1 forecasts would be to use statistical forecasts based on the history of the solar wind at L1. Here, we would then want to verify if the spectral behavior of solar wind velocity and magnetic field can be well reproduced through such empirical models. To facilitate such a study, we use the Analogue Ensemble model (AnEn) that is fast and easy-to-implement.  

AnEn has been introduced in Space Weather by \citeA{owens_probabilistic_2017} and \citeA{riley_forecasting_2017}, and builds an ensemble of forecasts from historical satellite measurements. The basic principle is that historical observations can provide a good analog, or ``similar day'', of the current conditions, and of the future. \citeA{lorenz_atmospheric_1969} proposed this method for meteorological forecasts and it is one of the paradigms used in neural network models \cite{burov_kernel_2020}. It is based on the idea that with a large enough dataset, we can find an ensemble of time periods in the past that display similar values and variations to current conditions. It is then assumed that the future evolution will be similar enough to make meaningful predictions. 
Such meaningful predictions could be used to fill observational data gaps that are pathological in space weather empirical geomagnetic models that couple the solar wind parameters to the geomagnetic indices. Following this aim, \citeA{lockwood_development_2019} analyzed the autocorrelation of various solar wind parameters at $1\unit{AU}$ and demonstrated that the reliability of persistence forecasts is inherently limited. Their study showed that while solar wind parameters exhibit high autocorrelation over short timescales (several hours), the correlation decays significantly over longer periods, reducing the accuracy of persistence-based forecasts for extended lead times, i.e. how long in the future, the forecast is accurate and can be used. One objective of this paper, is then to compare the AnEn forecasts to persistence, climatology and synodic recurrence baselines, to quantify its performance in terms of lead time accuracy and the associated relative predictability of the solar wind velocity and magnetic fields.  

\citeA{owens_probabilistic_2017} and \citeA{riley_forecasting_2017} applied the AnEn method to 1-hour near-Earth solar wind observations to forecast solar wind parameters (density, velocity, magnetic field). \citeA{haines_forecasting_2021} also applied this method to forecast geomagnetic indices. Empirically, these ensembles of forecasts contain all of the fluctuation information that has been measured, and hence the information to predict the solar wind Fourier spectra. However, that information can be lost when using bulk statistics of the ensemble (such as mean or median) for visualization, performance diagnostics and prediction purposes. Using spectral diagnostics, this paper tackles the question of the accuracy of this forecast method at higher resolution, the quantification of the loss of information in the fluctuations, and suggests an improved method to preserve the smaller-scale information in the analysis.


\section{Methods and data} \label{sec:method}
\subsection{The Analog Ensemble (AnEn) method} \label{subsec:AnEn}

\begin{figure}
    \centering
    \includegraphics[width=\linewidth]{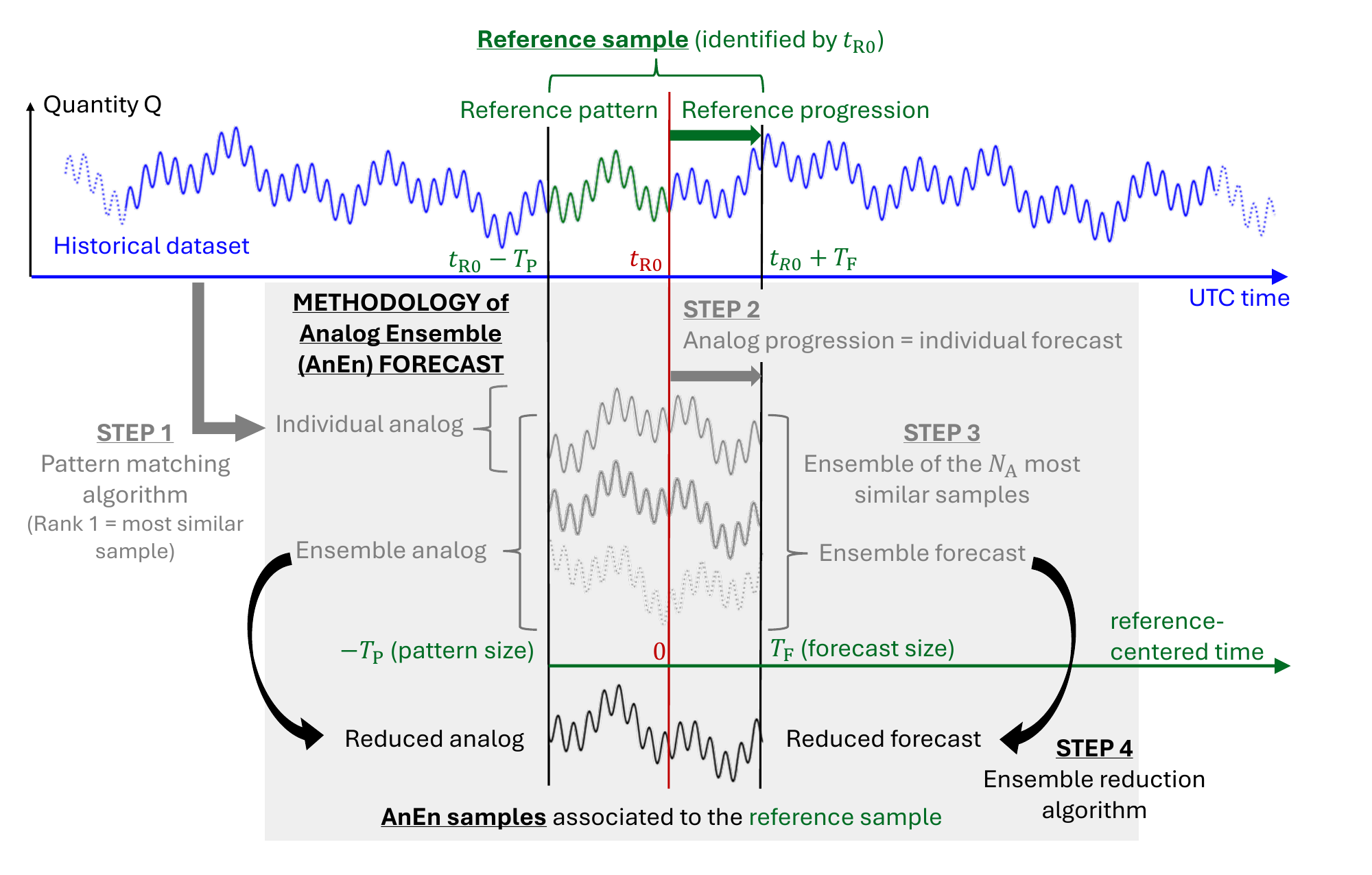}
    \caption{Summary of the AnEn method used in this article. The historical dataset (blue) provides individual analogs (STEP 1, grey, left side of the red line). Ensemble forecasts (STEP 3) are obtained from the progression of these analogs (STEP 2, grey, right side of the red line), and are reduced with STEP 4 (black). More details in section \ref{subsec:AnEn}.}
    \label{fig:AnEn_method}
\end{figure}

The analysis and forecast method follows the Analog Ensemble (AnEn) technique proposed by \citeA{owens_probabilistic_2017} and \citeA{riley_forecasting_2017}, which is schematically illustrated in Fig. \ref{fig:AnEn_method}. Consider a quantity $Q$ with the aim of making a forecast of the progression of $Q$ into the future, from the current, or ``reference'', time, $\tR$, to some maximum forecast lead time or ``forecast size'', $\TF$. 

We begin by defining a ``reference pattern'' formed from the recent observations of $Q$, spanning a time window from $\tR - \TP$ to the reference time $\tR$, where $\TP$ is referred to as the ``pattern size''. The rest of the dataset, i.e. prior to $\tR - \TP$ and after $\tR$, is used as a ``historical dataset'', from which analogs are drawn with the following procedure of pattern-matching. Each data point $\tA$ in this historical dataset is associated with an interval of length $\TP$, from $\tA - \TP$ to $\tA$, and each of these intervals, or ``individual analogs'', is compared with the reference pattern. The metric used to quantify the similarity between the reference pattern and the individual analogs is the mean square distance, 
\begin{equation}
\MSD = \left< (Q_{\text{individual analog}} - Q_{\text{reference pattern}})^2 \right>_{\TP}.
\end{equation}
We also require that individual analogs and reference patterns contain less than $\valueFLT{10}\unit{\%}$ of missing values in the computation of the $\MSD$. 
Minimizing the $\MSD$ allows us to attribute a rank to every individual analog from the most to the least similar to the reference pattern. A set of the $\NA$ most similar time series is extracted and forms the ``ensemble analog'' of rank $\NA$.

The progression of the time series after the end of an individual analog is the ``individual forecast''. Thus, an ``ensemble forecast'' of rank $\NA$ will be obtained from the progression of the individuals in the ensemble analog of rank $\NA$. While the ensemble forecast can be used as a probabilistic forecast, it is often reduced to a single time series through an ensemble reduction algorithm, such as computing the ensemble mean or median at each time step. The reduction of that ensemble to a single time series (as a prediction of the future behaviour) will be called the ``reduced forecast'' of rank equal to the rank of the least similar individual analog, $\NA$, considered in the ensemble analog. The duration of these time series will be noted $\TF$ for ``forecast size''. They will be compared to the progression of the reference pattern, after the reference time $\tR$, the ``reference progression''. The set of an analog and the associated forecast will be called the ``output sample'' while the set of the reference pattern and progression will be the ``reference sample''.

\subsection{Parameters of the statistical analysis}

A statistical analysis will be presented on the performance of output samples from $\NR=\valueFLT{200}$ reference samples and for several pattern sizes. $\NR=\valueFLT{200}$ is chosen for computational reasons. The pattern sizes $\TP$ are chosen from $\valueFLT{96}\unit{s}$ ($\valueFLT{4}~ \unit{ data points}$) to $\valueFLT{24}\unit{h}$ ($\valueFLT{3600}~ \unit{ data points}$). $\TP = \valueFLT{96}\unit{s}$ effectively tests if the progression depends on the most recent small-scale fluctuation, while $\TP = \valueFLT{24}\unit{h}$ contains the information from the previous days. However, note that the $\MSD$ will dilute small-scale fluctuations except for outlier values. An improvement of the forecast performance with a decrease of $\TP$ means that short-term past information is sufficient to model the future. A decrease in the performance of the forecast is expected with an increase of the forecast size $\TF$, as analogs increasingly diverge from the reference progression due to small initial differences growing with time and to unpredictable changes in the stream structure. Ensembles of up to $\NA = \valueFLT{2000}$ individuals will be considered. $\NA$ will also affect the performance. We expect the appearance of unpredictable events (ICME, stream variations, phase differences) in the ensemble forecast to be compensated when $\NA$ is large enough but that also means that progressions from less similar individual analogs will be accounted for.  

\subsection{Dataset}
The \textit{WIND} spacecraft \cite{acuna_global_1995} has been the only spacecraft at L1 providing sub-minute resolution measurements of the velocity and the magnetic field, resulting in long-term and high-resolution datasets \cite{wilson_iii_quarter_2021}. Its strategic position at L1 has been held since 2004. For computational reasons, the analysis is limited to the years from 2004 to 2009, corresponding to the decrease of activity of solar cycle 23 and the first year of solar cycle 24. Using L1 data to forecast L1 data restricts the analysis to solar wind historical occurrences in similar states of evolution, as its non-linear behavior evolves during its propagation in the interplanetary space  \cite{chen_evolution_2020}. The \textit{WIND} \textit{PLSP} dataset \cite{lin_wind_2021} provides ground computation of proton velocity amplitude and components in the geocentric solar ecliptic (GSE) coordinate system, resolved at $\valueFLT{24}\unit{s}$, based on measures of the 3DP (Three-Dimensional Plasma and Energetic Particle Investigation, \citeA{lin_three-dimensional_1995}) instrument, and synchronized with magnetic field amplitude and component measurements from MFI (Magnetic Field Instrument, \citeA{lepping_wind_1995}) instrument, also in the GSE coordinate system.
$\valueFLT{24}\unit{s}$-resolution provides most of the mesoscale or inertial range where no dispersion and dissipation occurs according to turbulence theories.
Due to large gaps in 2007 and 2008, 3 sub-datasets are considered -- from 3 May 2004 to 21 November 2006, from 1 October 2007 to 30 April 2008, from 1 June 2008 to 31 August 2009 -- in which anomalous values found in the velocity have been removed after comparison with \textit{WIND} SWE (Solar Wind Experiment, \citeA{ogilvie_swe_1995}) measurements resolved at $\valueFLT{94}\unit{s}$. One can note a slight shift in value affecting the velocity estimations of the PLSP dataset by comparison with the SWE one. As this shift varies slowly over multiple months that potentially affects the ranking of the analogs. This bias, due to the onboard calibration process, makes the analysis more realistic for a future application of the AnEn method for operational forecasting. The proportion of invalid values is estimated at $\sim\valueFLT{10}\unit{\%}$ for the first data subset, $\sim \valueFLT{27}\unit{\%}$ for the second subset and $\sim \valueFLT{35}\unit{\%}$ for the third subset. However, most gaps have a duration less than $\valueFLT{10}\ \unit{points}$ ($\valueFLT{4}\,\unit{min}$).  

$\valueFLT{495}$ potential reference times $\tR$ were selected carefully such that the error due to datagaps on the pattern matching algorithm (and hence on the performance diagnostics) are reduced and the ensemble is as representative as possible of the solar wind activity of 2004 and 2005. The requirements are:
\begin{itemize}
    \item no more than $ \valueFLT{5}\unit{\%}$ of missing values in patterns of several sizes $\TP$ log-regularly  chosen between $\valueFLT{96}\unit{s}$ ($\valueFLT{4}~ \unit{ data points}$) and $\valueFLT{24}\unit{h}$ ($\valueFLT{3600}~ \unit{ data points}$),
    \item no gaps larger than $ \valueFLT{4}\unit{\%}$ of the $\TP$, 
    \item no gaps in the progression larger than $\valueFLT{4}\unit{\%}$ of $\TF = 3600$ points,
    \item at least $ \valueFLT{12}\,\unit{h}$-interval  between two consecutive $\tR$.
\end{itemize}
Then $\NR = \valueFLT{200}$ reference times $\tR$ are randomly drawn from this ensemble of potential $\tR$. They are spread from 4 May 2004 03:26:24 to 24 December 2005 18:07:36. 

\subsection{Metrics and other forecasts used to diagnose the performances of the AnEn forecasts} \label{subsec:diag}

\begin{figure}
    \centering
    \includegraphics[width=\linewidth,trim={0cm 10cm 0cm 1cm},clip]{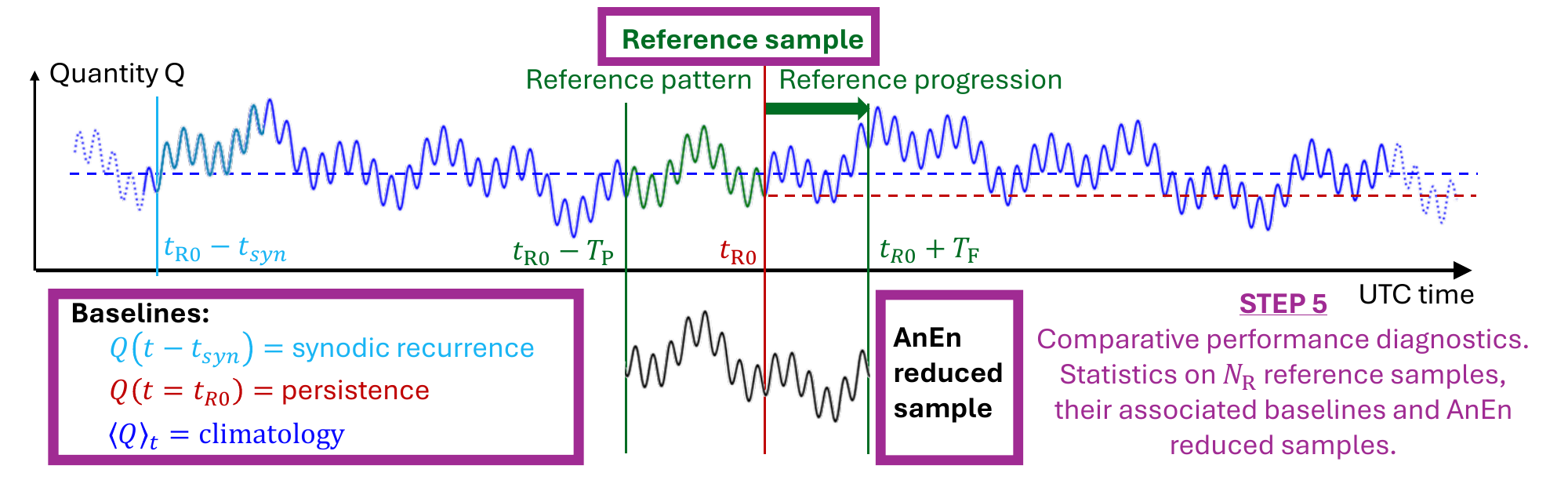}
    \caption{Summary of the progressions compared to quantify the performance of the AnEn forecasts. More details in section \ref{subsec:diag}.}
    \label{fig:diag}
\end{figure}

The performance of the AnEn forecast is computed by comparison to the observed variations (reference progression) and to several traditional forecasts. 
These are: a climatology baseline, wherein the future value of $Q$ is forecast to always be the historical average value, $\left<Q\right>_t$; a persistence baseline, wherein the future value of $Q$ is forecast to be equal to the value at the reference time, $Q(t=\tR)$; and a synodic cycle recurrence forecast, where the future value of $Q$ is forecast to be equal to that one solar synodic rotation period ($\tsyn$) earlier, $Q(t-\tsyn)$. We use $\tsyn=\valueFLT{27.125}\unit{days}$, following \citeA{owens_27_2013} that observed that the correlation peak is closer to $\valueFLT{27.125}\unit{days}$ for the velocity and the magnetic field main components and hence this value gives better forecasts than the synodic period of $\valueFLT{27.27}\unit{days}$.

To estimate the agreement between forecasts and observations, a number of metrics are considered. The first, associated to time-domain accuracy, is the Normalised Root-Mean-Square error: 
\begin{equation}
    \NRMSE = \sqrt{\frac{\left<(Q_{\text{forecast}}-Q_{\text{reference progression}})^2\right>_t}{\left<Q^2_{\text{reference progression}}\right>_t}}.
\end{equation}
The lower the $\NRMSE$, the better the forecast performance. By computing the AnEn $\NRMSE$ relative to that of a baseline forecast, it is possible to compute the AnEn skill: 
\begin{equation}
    \Skill = 1 -\frac{\NRMSE(\text{AnEn forecast})}{\NRMSE(\text{baseline})}
\end{equation}
that quantified the performance of an AnEn forecast relative to a baseline. If $\Skill > 0$, $\NRMSE(\text{AnEn forecast}) < \NRMSE(\text{baseline}) $, and the performance of the AnEn forecast is better than the performance of the baseline. 

The last metric is a spectral ratio that quantifies the scale-by-scale performance in the Fourier space:
\begin{equation}
    \SR(\freq) = \log_{10}\left(\frac{\|\FFT(Q_{\text{forecast}})\|}{\|\FFT(Q_{\text{reference progression}})\|}\right).
\end{equation}
It is fitted linearly from frequencies above $\decade{-4}\unit{Hz}$, assuming the frequency dependency is logarithmic:
\begin{equation}
\SR(\freq) \sim \fitSR = \slopeSR (\log_{10} \freq + 2) + \HFSR.
\end{equation}
Hence, it is possible to reduce the analysis of spectral performance at small scales to two parameters: the slope, $\slopeSR$, and the estimated spectral ratio at $\freq=\decade{-2}\unit{Hz}$, $\HFSR$. The sign and the value of $\HFSR$ indicate the gain or loss of power of the small-scale fluctuations. $\slopeSR$ indicates if this gain or loss is constant for all frequencies above $\decade{-4}\unit{Hz}$ or if the amplitude of the spectrum of $Q_{\text{output sample}}$ diverges from the one of $Q_{\text{reference sample}}$. The value $\decade{-4}\unit{Hz}$ is empirically chosen from the observed behavior of the particular case introduced in Fig. \ref{fig:3} and to give the same importance to all small-scale frequencies in the fit, $\SR(\freq)$ is binned log-regularly in frequency.


\section{Results}
The forecast of a particular reference time $\tR$, as a case study, is analyzed in Sections \ref{subsec:case_study} and \ref{subsec:case_study_diag}. The choice of algorithms for the reduction of the ensemble analogs, statistically compared in Section \ref{subsec:stat_NA}, is established from these results. 
They also illustrate the relationship between the ensemble analog paradigm and the baselines that will be used to define the predictability in Section \ref{subsec:predictability}.  


\subsection{Case study: 6 October 2004 14:58:00 UT and ensemble-reduction algorithm} \label{subsec:case_study}

\begin{figure}
\centering
\includegraphics[width=\textwidth,trim={1cm 1cm 1cm 0.5cm},clip]{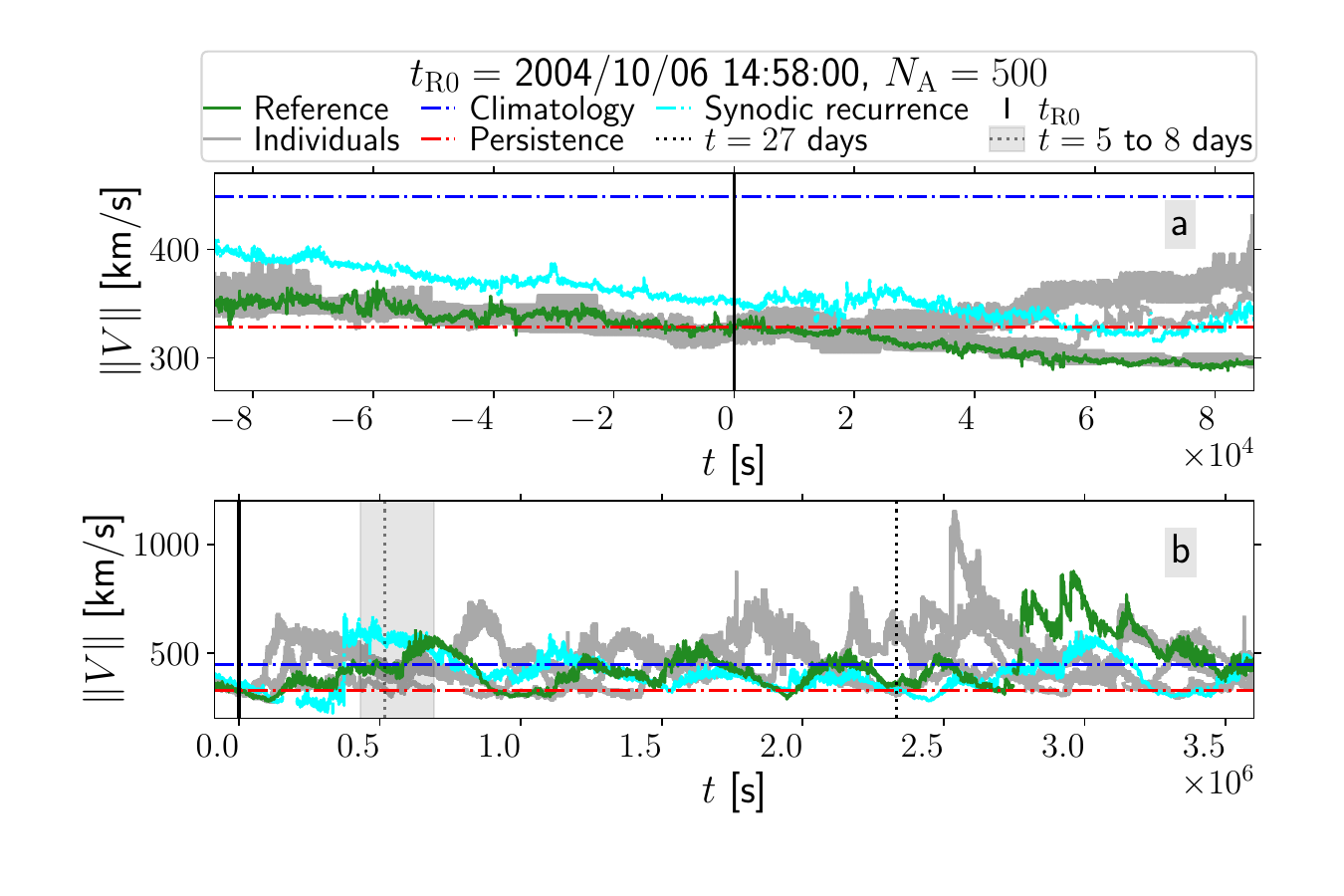}
\protect\caption{Forecasts of proton velocity magnitude for the reference time $\tR = $ 6 October 2004 14:58:00 UT (black vertical line). Short-term progression of size $\TF = \valueFLT{24}\unit{h}$ (a) and long-term progression $\TF = \valueFLT{42}\unit{days}$ (b) of the reference sample (green solid), $\NA = 500$ AnEn individual samples (grey solid) obtain with a pattern size $\TP= \valueFLT{24}\unit{h}$, the persistence (red dash-dotted), the climatology (blue dash-dotted), and the synodic rotation period recurrence (cyan dash-dotted). $\tR$ + 27 days (vertical dotted black) is the synodic rotation period,  $\tR$ + 5 to 8 days (dotted vertical grey line and area) is the period observed by Oloketuyi (2020).}
\label{fig:2a}
\end{figure}

Fig. \ref{fig:2a} is an example of forecasts for $\tR = $ 6 October 2004 14:58:00 UT with a pattern of size $\TP = \valueFLT{24}\unit{h}$ and two forecast sizes, $\TF = \valueFLT{24}\unit{h}$ (panel a) and $\TF = \valueFLT{1000}\unit{h} \sim \valueFLT{42}\unit{days}$ (panel b). The individual analogs (grey) obtained from applying STEP 1 and 2 of Fig. \ref{fig:AnEn_method} and the baseline forecasts, see Fig. \ref{fig:diag}, are displayed.
Panel a shows that while the individual analogs (grey) follow the reference pattern (green) quite closely, the individual forecasts (grey) tend to diverge from the reference progression (green) around 20000 seconds from $t=0$. This is around the same time as the persistence forecast (red dash-dotted) diverges from the reference progression. 
On longer time scales shown on panel b, both the reference progressions and individual forecasts vary around the climatological value (blue dash-dotted). 
This diverging behavior from a similar initial state is expected in a complex system following the rules of deterministic chaos \cite{yao_initial-condition_1995}, but on the longer timescales will result from changing conditions in the global solar wind structure too. A quasi-cyclic variation, or mean-reversibility, around the climatological value is also expected as the large-scale behavior of the solar wind follows multiple cycles. 
However, the synodic period of 27 days does not seem associated with this mean-reversibility. The first convergence to the climatology baseline seems to coincide with the first main period observed by \citeA{oloketuyi_responses_2020}, a period of 5 to 8 days, which was found in \textit{Wind} data during the time period of the present study using wavelet analysis. This periodicity is displayed with a grey vertical area (panel b). 
For this interval, the previous synodic period recurrence is not even in the first 500 analogs, too dissimilar to the reference pattern. However, long-term analysis reveals that it seems to reflect the main variations of the reference quite well.

\begin{figure}
\centering
\includegraphics[width=\textwidth,trim={1cm 1cm 1cm 0.5cm},clip]{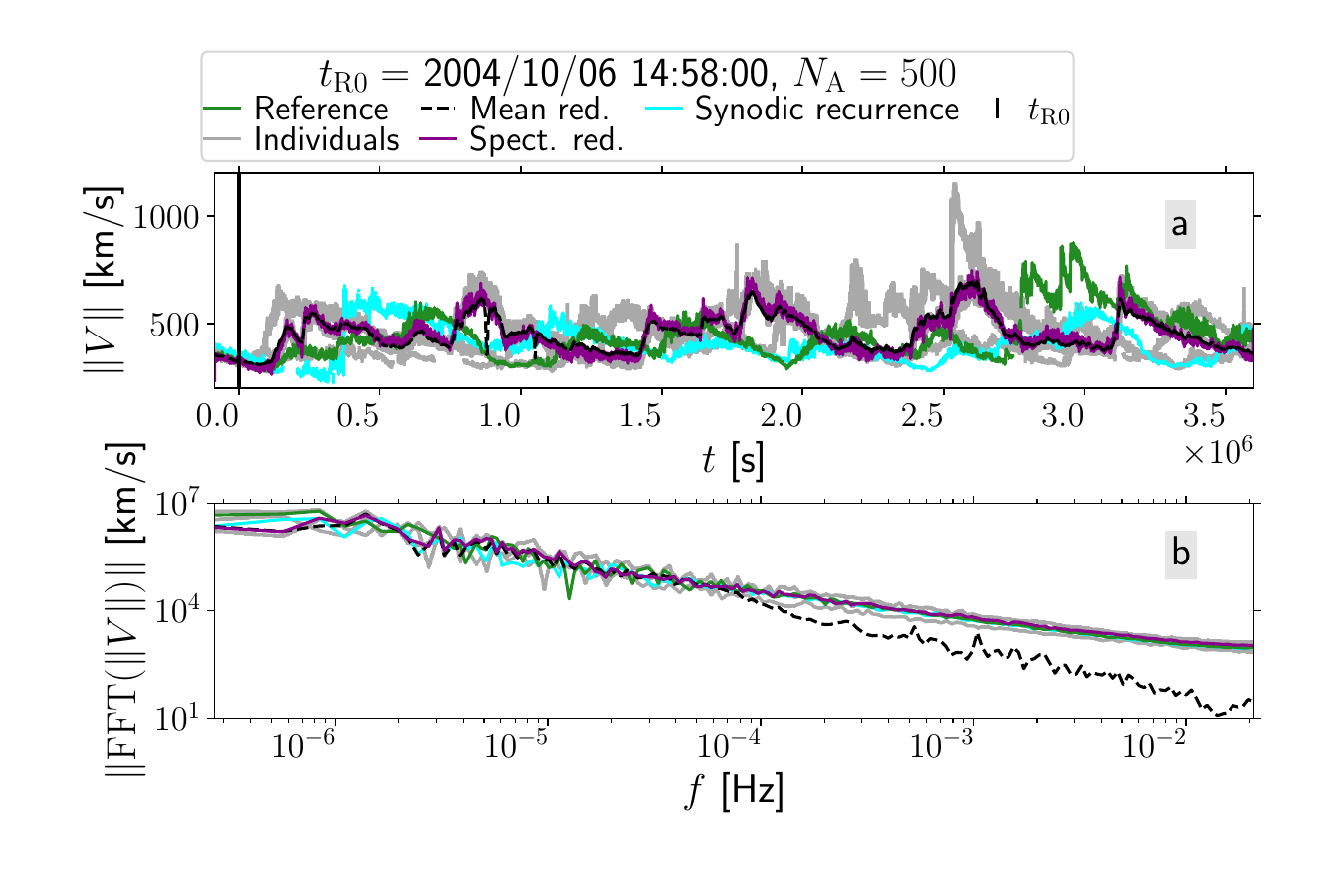}
\caption{Forecasts of proton velocity magnitude for $\tR = $ 6 October 2004 14:58:00 UT (black vertical line). Long-term progression $\TF = \valueFLT{42}\unit{days}$ (a) and spectra (b) of the reference sample (green solid), $\NA = 500$ AnEn individual (grey solid) obtain with a pattern size $\TP= \valueFLT{24}\unit{h}$, mean-reduced (dashed black dashed) and spectral-reduced (purple solid) samples, and the synodic rotation period recurrence (cyan dash-dotted). }
\label{fig:2b}
\end{figure}

\begin{table}
\centering
\caption{$\NRMSE$ performances and skill for the AnEn reduced samples and baselines of $\tR=$ 6 October 2004 14:58:00 UT with $\NA = 500$.}\label{tab:1}
\begin{tabular}{@{}lccc@{}}
\toprule
       Intervals    
       & $]-\TP;0]$ 
       & $]0;\TP]$ 
       & $]0;\TF]$ 
       \\ \midrule
\multicolumn{4}{c}{$\NRMSE$}   
\\ \hline
        Climatology              
        & 0.3151   
        & 0.4556                    
        & 0.2385           
        \\
        Persistence               
        & 0.0470   
        & 0.0756
        & 0.3428           
        \\
        Synodic recurrence               
        & 0.0991   
        & 0.1065
        & 0.2490           
        \\ 
        Mean red.         
        & 0.0162   
        & 0.0357             
        & 0.3090           
        \\
        Spect. red. to $\TP$         
        & 0.0199 
        & 0.0288             
        & N/A          
        \\
        Spect. red. to $\TF$         
        & 0.0305 
        & 0.0289            
        & 0.3073           
        \\\midrule
        \multicolumn{4}{c}{ 
            $\Skill$ over Climatology} 
        \\ \hline
        Mean red. 
        & 0.9487   
        & 0.9217        
        & -0.2956          
        \\
        Spect. red. to $\TP$      
        & 0.9370   
        & 0.9367           
        & N/A 
        \\
        Spect. red. to $\TF$      
        & 0.9031 
        & 0.9367         
        & -0.2883 
        \\ \midrule
        \multicolumn{4}{c}{
            $\Skill$ over Persistence} 
        \\ \hline
        Mean red. 
        & 0.6557   
        & 0.5281         
        & 0.0985          
        \\
        Spect. red. to $\TP$      
        & 0.5772
        & 0.6184       
        & N/A 
        \\
        Spect. red. to $\TF$      
        & 0.3495 
        & 0.6182         
        & 0.1036 
        \\\midrule
        \multicolumn{4}{c}{
            $\Skill$ over Synodic recurrence} 
        \\ \hline
        Mean red. 
        & 0.8370   
        & 0.6652         
        & -0.2410          
        \\
        Spect. red. to $\TP$      
        & 0.7997
        & 0.7293           
        & N/A 
        \\
        Spect. red. to $\TF$      
        & 0.6919  
        & 0.7292        
        & -0.2340
        \\
\bottomrule
\end{tabular}
\end{table}

\begin{figure}
    \centering
    \includegraphics[width=\linewidth,trim={1cm 4cm 2cm 4cm},clip]{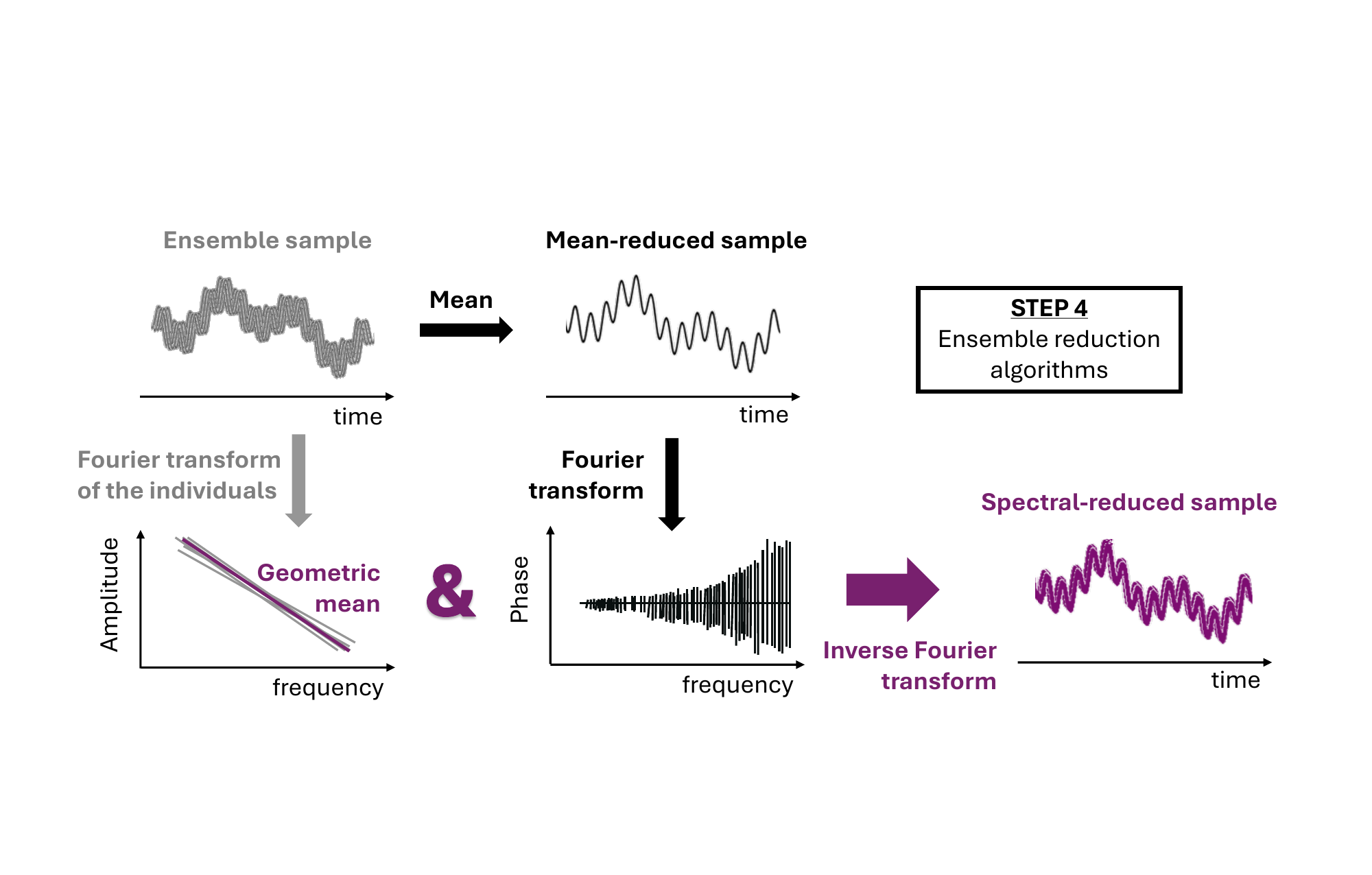}
    \caption{Summary of the reduction algorithm used to obtain the AnEn reduced forecast in STEP 4 of Fig. \ref{fig:AnEn_method}. More details in section \ref{subsec:case_study}.}
    \label{fig:reduct}
\end{figure}

Fig. \ref{fig:2b} shows the long-term, $\TF = \valueFLT{1000}\unit{h} \sim \valueFLT{42}\unit{days}$, time series (panel a) and spectra (panel b) of individual (grey) and synodic recurrence (cyan) samples along side reduced AnEn samples (STEP 4 of Fig. \ref{fig:AnEn_method}). Tab. \ref{tab:1} summarizes the time-accuracy $\NRMSE$ and $\Skill$ metrics of the baselines and the AnEn reduced analogs and forecasts. The first ensemble reduction algorithm applied is the traditional mean reduction (black dashed).
The amplitude of the spectra of the individual samples and the previous synodic recurrence follow the reference one. However, while the  AnEn mean-reduced forecast (black dashed line) seems to be quite accurate in terms of $\NRMSE$ by comparison with the baselines, the spectrum reveals a systematic loss of small-scale fluctuations (here shown for the interval $]-\TP;\TF]$ with $\TP= \valueFLT{24}\unit{h}$). Note that the frequency-by-frequency geometric average of the amplitude of the spectra of the individual samples (purple solid line) provides a closer level to the reference. 

Thus, we define a new reduction algorithm, here called ``spectral reduction" (Spect. red.). The reduced time series is produced following the algorithm of Fig. \ref{fig:reduct}. It uses the inverse Fourier transform of a composite spectrum whose amplitude is the geometric average of the amplitude of the spectra of the individual sample (purple spectrum), and the phase is the phase $\varphi_{\FFT}$ of the mean-reduced sample:
\begin{equation}
    \text{Spec. red.} = \invFFT\left(\sqrt[\uproot{5}\NA]{\prod_{\NA}\|\FFT(Q_{\text{individual sample}})\|} \cdot e^{i \varphi_{\FFT}\left(\left<Q_{\text{individual sample}}\right>_{\NA}\right)}\right) .
\end{equation}
The resulting time series is displayed in purple and the $\NRMSE$ and $\Skill$ measures associated are in the table (rows Spect. red.). The spectral reduction algorithm is based on the spectra of the whole samples instead of just the forecasts or the analogs. That ensures the coincidence at the reference time $\tR$ as edge effects are induced by the inverse Fourier transform of the composite spectra. In the following sections, the spectra used to do the reduction are computed on time series of forecast size equal to the analyzed interval. Indeed, for an interval $]0;\TP]$, the reduction will be applied on samples of size $]-\TP;\TP]$ even if using a larger sample does seem to only affect slightly the $\Skill$ performance in Table \ref{tab:1} (comparison between rows Spect. red. and column $]0;\TP]$). 

\subsection{Quantification of the small scale performance with the case study: 6 October 2004 14:58:00} \label{subsec:case_study_diag}

\begin{figure}
\centering
\includegraphics[width=\textwidth,trim={1cm 1cm 0 0},clip]{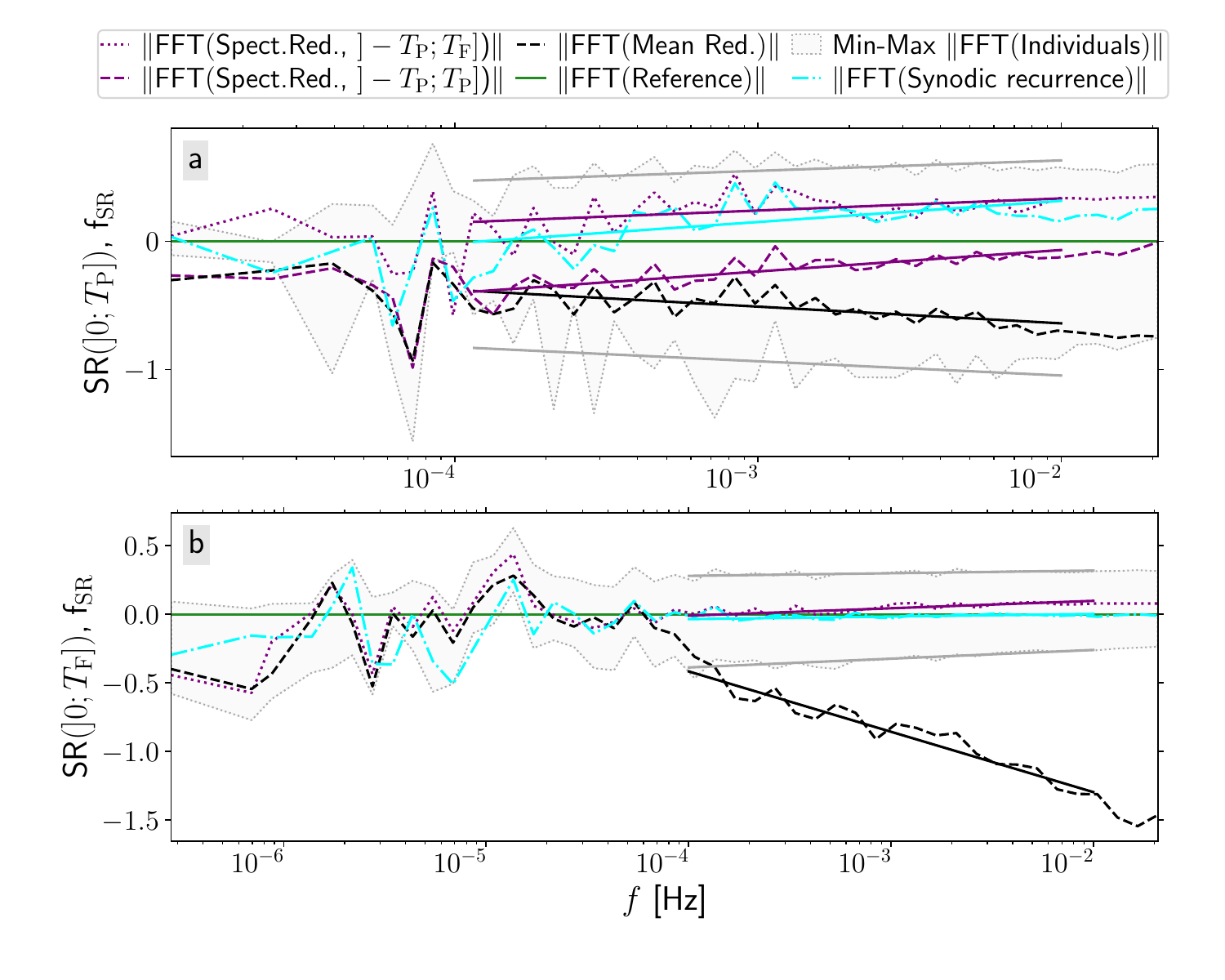}
\caption{Spectral ratio of short-term (panel a) and long-term (panel b) forecasts and respective small-scale fits (straight lines). Green: $0$-value, the threshold between over- and under-estimation. Cyan: spectral ratio $\SR$ applied to the spectral amplitude of the synodic recurrence forecast. Grey: range between the minimum and the maximum spectra obtained from the AnEn ensemble forecast. Black: $\SR$ applied on the spectra of the AnEn mean-reduced forecast. Purple: $\SR$ applied on the spectra of the spectral-reduced forecasts obtained for AnEn individual samples of size $]-\TP;\TP]$ (dashed) or $]-\TP;\TF]$ (dotted). The results are binned on log-regular 50 frequencies for readability. Example of the proton velocity magnitude.}
\label{fig:3}
\end{figure}

\begin{table}
{\centering
\caption{ Slope, $\slopeSR$, and values at $\decade{-2}\unit{Hz}$, $\HFSR$, associated to the spectral ratio fits $\fitSR$ of Fig. \ref{fig:3}}\label{tab:2}
\begin{tabular}{@{}r|cc|cc@{}}
\toprule
       & \multicolumn{2}{c}{$\SR(]0,\TP])$} 
       & \multicolumn{2}{c}{$\SR(]0,\TF])$}
       \\ \hline
       Type
       & $\HFSR$
       & $\slopeSR$
       & $\HFSR$
       & $\slopeSR$
       \\ \hline
       Mean red.
       & $-0.643$
       & $-0.132$ 
       & $-1.294$
       & $-0.453$ 
       \\
       Spect. red. to $\TP$  
       & $-0.069$ 
       & $0.166$ 
       & N/A
       & N/A
       \\
       Spect. red. to $\TF$  
       & $0.363$ 
       & $0.111$ 
       & $0.101$ 
       & $0.055$ 
       \\
       $\text{Min}_{\NA}~ \|\FFT(\text{individuals})\|$
       & $-1.022$ 
       & $-0.087$ 
       & $-0.248$ 
       & $-0.071$ 
       \\
       $\text{Max}_{\NA}~ \|\FFT(\text{individuals})\|$
       & $0.628$ 
       & $0.078$ 
       & $0.321$ 
       & $-0.020$ 
       \\
       \bottomrule
\end{tabular}}
\\
{\textit{Note.} Short-term $]0,\TP]$ and long-term $]0,\TF]$ AnEn forecasts of $\tR=$ 6 October 2004 14:58:00 with $\NA = 500$ individual forecasts, a pattern size $\TP = \valueFLT{24}\unit{h}$ and a maximal forecast size $\TF = \valueFLT{42}\unit{days}$}.
\end{table}

Fig. \ref{fig:3} illustrates the quantification of the spectral performances with the spectral ratio $\SR$ applied on short-term (panel a) and long-term (panel b) forecasts. A value of 0 corresponds to the ideal case when the forecasted spectral amplitude is the same as the reference, a value greater than $0$ indicates an overestimation, and a value less than $0$ indicates an underestimation (equivalent to a loss of power). At all spectral scales for short-term progression and large scales for long-term progression, the AnEn mean reduction (black), the spectral reduction (purple) and the synodic recurrence (cyan) give similar results. These forecasts fall within the span of the minimum and maximum values of the spectral amplitude of the ensemble forecast (grey area). For all forecast times, this span is centered on the $0$-value, meaning that the ensemble forecast reflects well the spectral properties of the reference progression.

However, the AnEn mean-reduced forecast diverges to smaller values of the spectral ratio $\SR$ at small scales, slightly for short-term progression and more obviously for long-term. This divergence is quantified in Tab. \ref{tab:2} with the slope $\slopeSR$ and the value at $\decade{-2}\unit{Hz}$ $\HFSR$ of the small-scale fits. It happens for frequencies higher than $\decade{-4}\unit{Hz}$, with a break-point between $\decade{-5}\unit{Hz}$ and $\decade{-4}\unit{Hz}$. This is also approximately the value at which the AnEn begins to diverge from the persistence forecast. This could be a signature of phase shifts between individual samples at small scales, whereas similar phase shifts at the largest scales do not result in a loss of power in the associated fluctuations. As illustrated by Fig. \ref{fig:2a}, most individual samples are consecutive shifts (small-scale shifts whose maximum is defined by the correlation time $\corrtime =  1.2\times10^{5}\unit{s}$, a diagnostic of the auto-similarity of the time series) of the higher-ranked one and have superposed spectra. Considering a large AnEn size $\NA$ seems to have effectively randomized the ensemble distribution of smaller-scale fluctuations, and flatten their amplitude through the mean reduction, while the spectral reduction preserves them. The grey area, larger for short-term progression, reveals a statistical convergence of the spectra of the individual forecasts -- statistics on the duration of the time series. In spectral and other turbulence statistical investigation, this duration is usually compromised with the number of individuals (that here would be equivalent to $\NA$, and the purple long-term results) \cite{isaacs_systematic_2015}.


\subsection{Effect of the rank of the ensemble forecast on statistical performances over 200 reference samples} \label{subsec:stat_NA}

\begin{figure}
    \centering
\includegraphics[width=\linewidth]{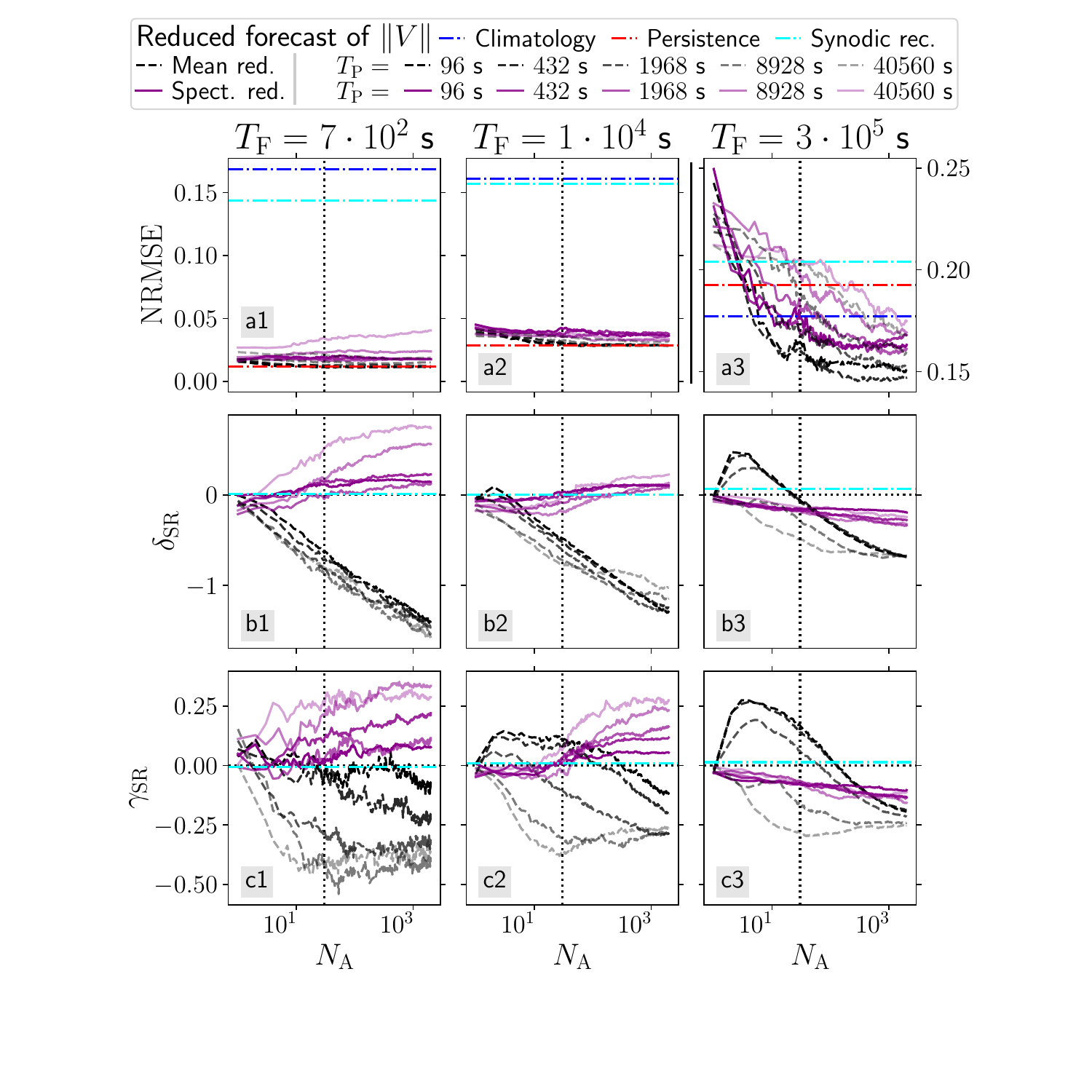}
    \caption{Summary of statistical performances (median on $\NR=200$ reference times) of AnEn mean-reduced (dashed black) and spectral-reduced (purple) forecasts, according to the forecast duration $\TF$ (columns), the pattern size used for the forecasts $\TP$ (lightness) and the size of the forecast ensemble $\NA$ (horizontal axis). Example for proton velocity magnitude.  
    Panels a1 to a3: the time-accuracy $\NRMSE$ better if minimal. Panels b1 to b3: the high frequency level $\HFSR$ better close to $0$ (dotted black horizontal line). Panels c1 to c3: the slope $\slopeSR$ better close to $0$ (dotted black horizontal line). Dotted vertical line: value $\NA=30$ selected for Sec. \ref{subsec:predictability}. Dash-dotted lines: baseline forecasts.}
    \label{fig:4}
\end{figure}

The forecast performance is now assessed by computing median values of the time-accuracy metric $\NRMSE$, and the frequency-accuracy metrics $\HFSR$, and $\slopeSR$. The metrics are applied on AnEn reduced forecasts computed for three forecast sizes and five pattern sizes. In order to limit computational cost, the AnEn forecasts are estimated at $\NR = 200$ reference samples.  Fig. \ref{fig:4} summarizes this study according to the rank of the ensemble forecast, $\NA$ i.e. the rank of similarity of the less similar analog used to obtain the ensemble, from 1 to 2000, for the proton velocity magnitude (similar behaviors have been found for the magnetic field magnitude and their components).

For the $\NRMSE$ (panels a1 to a3), mean-reduced forecast performance is, on average, better than spectral-reduced. So the large fluctuations brought by the spectral correction seem to worsen this accuracy metric. As expected, the accuracy increases with the size of the ensemble; a larger ensemble size means a smoothing of extreme events (coronal mass ejection, etc.). Even with $\NA = 2000$ individuals in the ensembles, we are still not able to include sufficiently dissimilar and varying individuals in the AnEn ensembles due to the high resolution of the dataset limiting the computational load. The largest pattern size and smaller forecast size make an exception: as the values closer to the reference time $\tR$ are lost in the $\MSD$, chosen as recognition criteria, the AnEn ensembles can include more dissimilar individuals. For the forecast sizes $\TF = 7\times10^2\unit{s}$ and $1\times10^4\unit{s}$, the performances are close to that of persistence, while for $3\times10^5\unit{s}$ they are better than all other models (a more rigorous study according to $\TF$ is proposed in Section \ref{subsec:predictability}).

$\HFSR$ (panels b1 to b3) confirms a general underestimation of small-scale power for the mean-reduced forecast, except for small ensemble size $\NA$ and small forecast size $\TP$, most likely due to a lack of statistics in the ensemble forecasts to effectively smooth the small-scale fluctuations (similarly to the $\TF = \TP$  case of Sec. \ref{subsec:case_study}). The slope $\slopeSR$ (panels c1 to c3) is mostly negative too, indicating a divergence of the AnEn mean-reduced forecast spectral amplitude from the spectra of the reference progression. For small pattern sizes, $\slopeSR$ can even be positive (divergence is the high-frequency value is positive/overestimation, convergence in the other scenario) but still close to zero, indicating that the spectra are more or less parallel. $\HFSR$ and $\slopeSR$ of the spectral-reduced forecasts closer to 0 than the mean-reduced ones prove the effectiveness of the spectral reduction. However, a divergence from zero is still observed with the increase of $\NA$.

Consequently, the choice of an optimal ensemble size $\NA$ has to be a balance between absolute accuracy (low $\NRMSE$) and minimising loss of fluctuation power (values of $\slopeSR$ and $\HFSR$ close to zero), whatever the quantity and the choice of reduction method. $\NA=30$ seems to be a good compromise for all quantities. \citeA{riley_forecasting_2017} chose $\NA = 50$,  while \citeA{owens_probabilistic_2017} chose $\NA = 100$, compromising the accuracy with numerical computation reasons. Such values would also work quite well for our dataset resolved at $24\unit{s}$ instead of $1\unit{h}$ according to Fig. \ref{fig:4}.


\subsection{Predictability and statistical performances for the velocity and magnetic field} \label{subsec:predictability}

\begin{figure}
\centering
\includegraphics[width=\textwidth]
{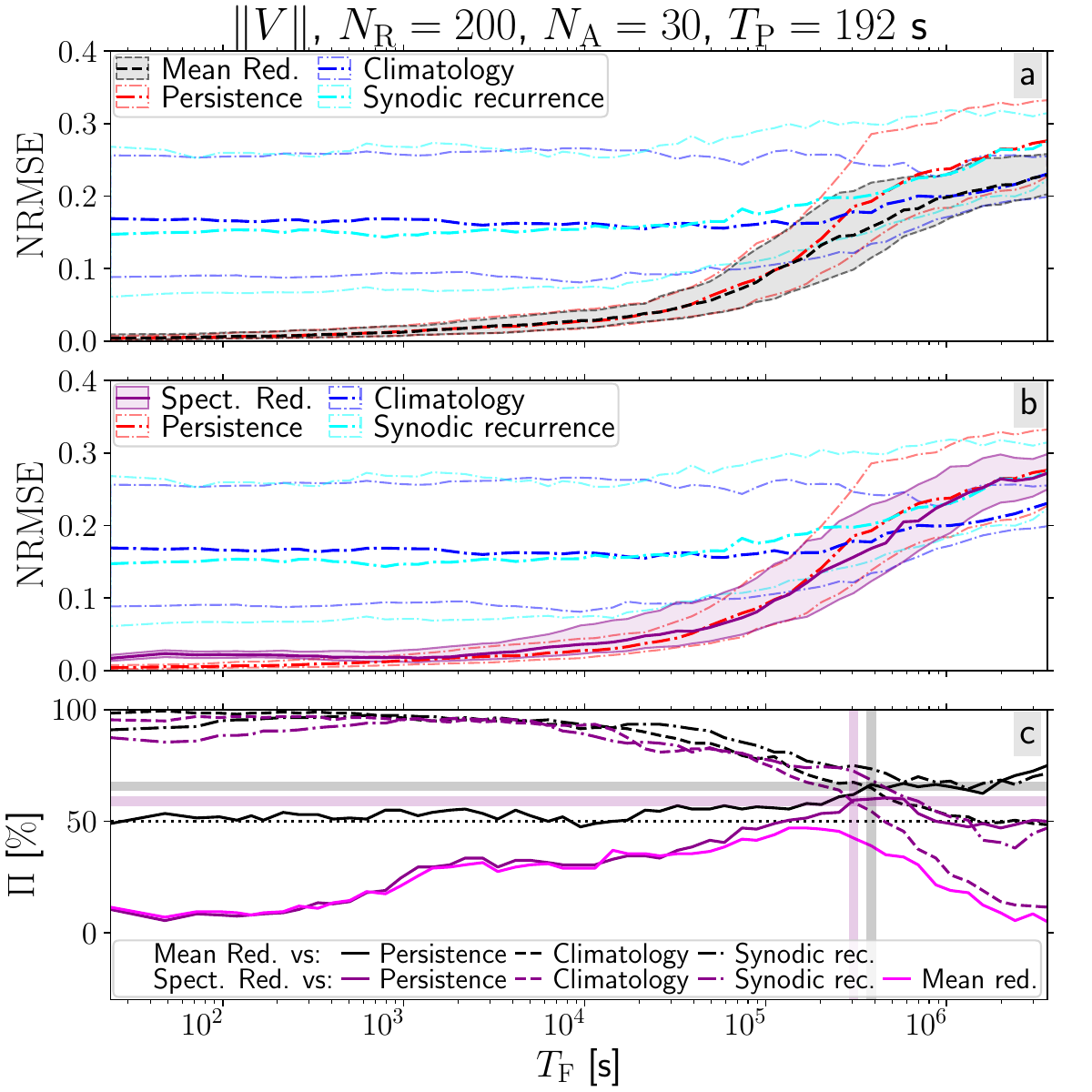}
\caption{Performance of mean-reduced (black, panels a and c) and spectral-reduced (purple, panels b and c) forecasts of proton velocity magnitude as a function of the size of the forecast, $\TF$. Results are shown for $\NA=30$ individuals, $\NR=200$ reference times and a pattern size $\TP=192\unit{s}$. 
Panels a and b: Median and quartiles (lighter) of $\NRMSE$. Comparison with the persistence (red dash-dotted), the climatology (blue dash-dotted), and the synodic recurrence (cyan dash-dotted).
Panel c: Proportion of positive $\Skill$ relative to persistence (straight), to climatology (dashed), and to synodic recurrence (dash-dotted), and of spectral-reduced relative to mean-reduced forecasts (magenta). The crosses indicate the predictability $\predict = \#_{\NR}(\Skill>0)$ where $\predict_{persistence} = \predict_{climatology}$ and the associated optimal lead time $\TL$ for mean-reduced (grey) and spectral-reduced (pink) forecasts.}
\label{fig:5}
\end{figure}

\begin{table}
{\centering
\caption{Summary of predictability and performances of AnEn mean- and spectral-reduced forecasts and synodic recurrence for the velocity and the magnetic fields}\label{tab:3}
\begin{adjustbox}{width=1\textwidth}
\begin{tabular}{@{}lcccccccc@{}}
\toprule    
       & $\|V\|$ 
       & $V_x$ 
       & $V_y$
       & $V_z$
       & $\|B\|$ 
       & $B_x$ 
       & $B_y$
       & $B_z$
       \\ \midrule
\multicolumn{9}{l}{$\predict~ [\unit{\%}]$}   
\\ \hline
        Mean red.               
        & 66
        & 65
        & 61
        & 64
        & 58
        & 66
        & 63
        & 63
        \\
        Spect. red.               
        & 59
        & 58
        & 56
        & 52
        & 47
        & 56
        & 54
        & 53         
        \\\midrule
\multicolumn{9}{l}{$\TL~ [\unit{s}]$} 
        \\ \hline
        Mean red. 
        & $\valueSI{3.8}{5}$
        & $\valueSI{3.8}{5}$
        & $\valueSI{3.9}{4}$
        & $\valueSI{3.3}{4}$
        & $\valueSI{1.7}{5}$
        & $\valueSI{1.3}{5}$
        & $\valueSI{2.5}{5}$
        & $\valueSI{1.4}{4}$
        \\
        Spect. red.     
        & $\valueSI{3.1}{5}$
        & $\valueSI{3.1}{5}$
        & $\valueSI{3.3}{4}$
        & $\valueSI{3.9}{4}$
        & $\valueSI{1.7}{5}$
        & $\valueSI{1.7}{5}$
        & $\valueSI{2.5}{5}$
        & $\valueSI{0.7}{4}$
        \\
        $\tau_c~ [\unit{s}]$     
        & $\valueSI{1.2}{5}$
        & $\valueSI{1.2}{5}$
        & $\valueSI{1.4}{4}$
        & $\valueSI{1.5}{4}$
        & $\valueSI{0.4}{5}$
        & $\valueSI{0.4}{5}$
        & $\valueSI{0.3}{5}$
        & $\valueSI{0.4}{4}$
        \\
        $\lambda_c~ [\unit{km}]$     
        & $\valueSI{6.3}{7}$
        & $\valueSI{6.3}{7}$
        & $\valueSI{6.8}{6}$
        & $\valueSI{7.4}{6}$
        & $\valueSI{2.2}{7}$
        & $\valueSI{2.3}{7}$
        & $\valueSI{1.7}{7}$
        & $\valueSI{2.1}{6}$
        \\ \midrule
\multicolumn{9}{l}{
            $\NRMSE$} 
        \\ \hline
        Mean red.               
        & 0.152
        & 0.153
        & 0.747
        & 0.940
        & 0.317
        & 0.934
        & 0.957
        & 0.918
        \\
        Spect. red.               
        & 0.165
        & 0.159
        & 0.804
        & 1.017
        & 0.358
        & 0.965
        & 1.020
        & 0.963
        \\
        Synodic rec.             
        & 0.199
        & 0.199
        & 1.197
        & 1.348
        & 0.461
        & 1.211
        & 1.149
        & 1.390
        \\\midrule
\multicolumn{9}{l}{
            $\HFSR$} 
        \\ \hline
        Mean red.               
        & $-0.048$
        & $-0.053$
        & $-0.485$
        & $-0.513$
        & $-0.220$
        & $-0.457$
        & $-0.484$
        & $-0.484$
        \\
        Spect. red.               
        & $-0.145$
        & $-0.137$
        & $-0.163$
        & $-0.163$
        & $-0.118$
        & $-0.201$
        & $-0.209$
        & $-0.210$
        \\
        Synodic rec.            
        & $0.041$
        & $0.040$
        & $-0.011$
        & $-0.002$
        & $0.002$
        & $0.023$
        & $0.016$
        & $0.020$
        \\
\bottomrule
\end{tabular}
\end{adjustbox}}
{\textit{Note.} AnEn best predictability $\predict$ and associated forecast size, i.e. optimal lead time $\TL$, for mean-reduced and spectral-reduced forecasts of the magnitude and components of the velocity and the magnetic fields. The lead times correspond to several autocorrelation times $\corrtime$ (associated with autocorrelation lengths $\corrlength$). Spectral reduction offers a compromise between time series accuracy (median of $\NRMSE$ at $\TF = \TL$ better than the one of the synodic recurrence baseline) and the loss of high-frequency fluctuations (median of $\HFSR$ at $\TF=\TL$ better than one of mean-reduced forecasts, except for $\|V\|$ and $V_x$ in accordance with Fig. \ref{fig:4}). Values obtained for $\NA = 30$ individuals, the pattern size $\TP = \valueFLT{192} \unit{s}$ and statistics over $\NR = 200$ reference times.}
\end{table}

One crucial question is the predictability of future solar wind evolution. Two quantities are of interest: how many intervals of the solar wind can be predicted? and how long in the future can we forecast?
To investigate this, we use AnEn reduced forecasts of rank $\NA = 30$ obtained for a pattern size $\TP = 192\unit{s}$ and associated with $\NR = 200$ references, parameters chosen from the results of Sec. \ref{subsec:stat_NA}. Sec. \ref{subsec:stat_NA} reveals that small values of $\TP$ are better, and similar performance for $\TP = 96 \unit{s}$ and $\TP = 432 \unit{s}$. For these parameters, the maximal dissimilarity of the considered analogs ($\sqrt{\MSD}_{\NA=30}$) is statistically $1\unit{km/s}$ for the magnitude and the components of the velocity and $0.1\unit{nT}$ for the magnetic field.
The predictability, $\predict$, is defined as the proportion of reference samples better forecast by AnEn than the baseline. It is contained in the positive $\Skill$ diagnostic depending on the ratio between a reduced forecast and the baseline forecast. $\predict>50\%$ means that the reduced forecast performs better for more than half of reference samples, a sign of a good predictability relative to the $\NRMSE$ diagnostic. The forecast size associated to the best predictability, $\TL$, corresponds to the optimal lead time, characteristic of the progression duration such that the AnEn is the optimal choice of forecast model (relative to persistence and climatology). The optimal lead time for the persistence baseline for instance, is the correlation time $\corrtime$ and its spatial equivalent the correlation length $\corrlength$ (estimations are provided in Table \ref{tab:3}). 

The $\NRMSE$ for each forecast type (panels a and b) and the proportion of positive skill $\predict$ (panel c) for different forecasts are displayed in Fig. \ref{fig:5} for the proton velocity magnitude. The performance of the AnEn mean-reduced forecasts follows the persistence baseline for the forecast sizes $\TF < 2\times10^5\unit{s}$ and then converges to the climatology. This behavior has been observed by \citeA{owens_probabilistic_2017}. The spectral reduced forecasts diverge slightly from the AnEn mean reduced forecast at small and large $\TF$. At large $\TF$, it seems to converge to the performance of either the synodic cycle forecast or persistence. However, the latter behavior varies according to the investigated quantity. As expected, the distribution of the $\NRMSE$ accuracy of the $\NR$ AnEn reduced forecasts, revealed by the inter-quartiles ranges surrounding the median in Fig. \ref{fig:5}a and b, is converging at small $\TF$ similarly to the persistence inter-quartile. It diverges slightly at larger $\TF$ due to the divergence of the forecast from the reference progression. The latter is within the climatology inter-quartile range. The synodic cycle forecast for the velocity magnitude is similar to climatology according to the median of the $\NRMSE$ for small $\TF$, however the quartiles indicate that the distribution is very broad and slightly larger for the synodic cycle. Depending on the historical dataset and not minimizing the error with the reference last point value, they are less relevant than the persistence that is based on the last value of the pattern reference. For $\Skill$ (panel c), the mean reduced and the spectral reduced forecasts can perform better than the three baselines with optimal values $\predict$ and $\TL$ referenced in Tab. \ref{tab:3}.

The table of Fig. \ref{tab:3} summarizes the predictability $\predict$ and the optimal lead time $\TL$ for every quantity. The magnetic field and velocities are on average predictable at $60\unit{\%}$ with AnEn, with slightly better results for AnEn mean-reduced forecasts. The main difference between the quantities is the optimal lead time that corresponds to several autocorrelation time, $\corrtime$. This optimal lead time is similar for both reduction algorithms. $\corrtime$ has been computed on chunks smaller than $22\unit{days}$ over the whole the historical dataset, and correspond to the smallest scale such that the normalized autocorrelation has dropped below $1/e$. For $\|V\|$, $V_x$, $\|B\|$, $B_x$, and $B_y$, the lead time is around 2 to 3 days, while $V_y$, $V_z$, and $B_z$ give the worst results of lead times, around several hours. As observed previously, the best performance of AnEn happened when the persistence performance are decreasing in favor of the climatology. As $V_y$, $V_z$ and $B_z$ are not the main components of their field and contain primarily fluctuations, their last pattern value is not expected to last long in the future. Indeed, $V_x$ contains most of the magnitude as the flow is predominantly radial direction, while $\mathbf{B}$ is mainly in the ecliptic plane $(x,y)$ but not radial direction due to the Parker spiral. These discrepancies agree with auto-correlations studies implying that the values of $B_z$ persist less than the velocity and, hence, that this quantity is harder to predict \cite{lockwood_development_2019,wicks_variation_2010}. The estimated values of correlation length, $\corrlength$, obtained by multiplying the estimation of $\corrtime$ associated to each chunk by its average proton velocity magnitude, are 10 times above the estimation of \citeA{wicks_variation_2010} done for 2004 to 2009 with L1 observations resolved at $1\unit{min}$. However, such estimations vary heavily according to the size of the considered chunks (closer values have been obtained by using accordingly $1\unit{day}$ intervals instead of $22\unit{days}$). 

The time-accuracy $\NRMSE$ and frequency-accuracy $\HFSR$ applied on the AnEn mean-reduced and spectral-reduced forecasts and the synodic recurrence with $\TF = \TL$ are also provided in Tab. \ref{tab:3}. For all quantities, the spectral reduction provides time accuracy slightly worse than the mean reduction but not as much as the synodic recurrence. 
For $V_y$, $V_z$, $\|B\|$, $B_x$, $B_y$, and $B_z$, the frequency accuracy is improved relative to the mean-reduced forecasts. $\|V\|$ and $V_x$ provide the opposite behavior as the optimal lead time for the chosen parameters, $\NA = 30$ individuals and the pattern size $\TP = 192\unit{s}$, is close to the $\TF$ investigated in Fig. \ref{fig:4}b3. For all quantities, the chosen parameters do not provide a better frequency accuracy than the synodic recurrence. However, these values provide insight on how the AnEn spectral-reduced forecast compromised the time and frequency accuracies.


\section{Discussion and conclusion}

We have presented a study of the analog ensemble (AnEn) method (which is based on pattern matching of past behaviour of a system to make predictions about its future), with a new emphasis on its ability to capture the multi-scale nature of the solar wind accurately. It is one of the few space weather modelling tools that can provide ensemble forecasts capturing and predicting the solar wind mesoscale variations. These scales are important for estimating the geomagnetic response to solar activity \cite{owens_ensemble_2014,ala-lahti_impact_2024}.

In this study, we have demonstrated several techniques that can be applied for performance estimation and operational forecasts. The performance of the AnEn method with varying ensemble size and forecast duration is quantified using a high-resolution dataset ($24\unit{s}$). A spectral performance diagnostic is introduced to investigate the small-scale statistical accuracy of the forecasts. A new way for reducing the ensemble of forecasts to a future prediction of the time series is introduced through a spectral reduction method that (unlike previous methods) preserves small-scale fluctuation amplitudes.

The AnEn method is found to provide forecasts of performance close to the persistence at short-term progression and the climatology at long-term, but better than both at intermediate progression lengths. It can also, according to the desired aspect (for instance, for short-term forecast and time-accuracy), be significantly better than the synodic cycle forecast. Statistically, the AnEn provides better results in about $60\%$ of the reference intervals of the velocity and the magnetic field-related quantities than persistence and climatology. That enhanced predictability is mostly reached for lead times associated with transitions from persistence-like to climatology-reversible durations. For the main components of the velocity ($Vx$) and the magnetic field ($Bx$ and $By$), the lead time is around 2 to 3 days, and around several hours for the other components agreeing with correlation-time analyses done in L1. While large scales and the short-term are better forecasted with AnEn, the synodic recurrence gives a more accurate forecast of small-scale variations in the Fourier space. However, AnEn spectral reduced forecasts provide a better compromise between time and frequency accuracy than the mean-reduced forecasts, maximising the time-accuracy, or the synodic recurrence, maximizing the frequency accuracy.

In addition, there are some caveats to consider regarding this study. The first concerns data gaps or missing values, which can affect the reduction algorithm output. For the mean reduction, the missing values are omitted. However, before applying the Fourier transform needed by the spectral reduction, the missing values in the samples must be linearly interpolated. This results in the discrepancies between the spectral reduced and mean reduced forecast performances of Fig. \ref{fig:4} at small ensemble size $\NA$. In the ideal case of no missing values, the performance would be the same for $\NA=1$. They converge for spectral performances as the diagnostic also interpolates the missing values of the mean reduced forecasts. Data gaps also affect the pattern matching algorithm. To mitigate their effect on the forecast performance analysis, several precautions have been applied, as described in Sec. \ref{sec:method}. However, missing values are the reality of real-time forecasts. So a rigorous investigation of their effects on the AnEn forecasts would be needed to support a proof of concept of this method for operational applications. Such an analysis has been proposed for persistence and auto-correlation by \citeA{lockwood_development_2019}.

The second caveat is the historical dataset. Here, around four years are used, spanning the transition from the solar cycle 23 to the solar cycle 24. Thus, we test the usefulness of the AnEn methodology during low solar activity period. A comparison with high solar activity period, such as the cycle 25 that provided the 10 May 2024 geomagnetic storm is left for future studies. Operational applications would use all available historical dataset, and potentially pattern matching over multiple quantities and/or spacecraft. An example of pattern-matching algorithm using multiple quantities is given by \citeA{owens_probabilistic_2017}. Other biases associated with the historical dataset come from its quality, in terms of calibration. However, not trying to compensate the miscalibrations in the PLSP dataset that have surely affected the pattern matching ranking is more realistic for operational purposes: what forecast can we provide with the data at our disposition?    

The final caveat is about the use of the mean over the median. The $\MSD$ criterion is fast, but it loses significance for large pattern sizes, as more potentially irrelevant information is considered, and it is sensible to extreme values due to random localized events or instrumental nonphysical fluctuations. \citeA{owens_probabilistic_2017} and \citeA{riley_forecasting_2017} used a median reduction algorithm to build the reduced forecast. 

To conclude, the AnEn methodology can provide forecasts that include small-scale fluctuations, which could enhance solar wind physics-based models or data assimilation methods \cite{lang_improving_2021}by providing well-forecasted data to assimilate. It also provides more accurate upstream conditions to inject into magnetospheric response models, for instance, complementing large scale physics-based forecasts through downscaling \cite{owens_ensemble_2014}. This work investigates the balance between the time- and frequency-domain accuracy associated with the AnEn reduction methods and the interpretation of such ensemble forecasts.


\section*{Acknowledgments}
P. A. Simon, C. H. K. Chen, and C. Sishtla are supported by UKRI Future Leaders Fellowship MR/W007657/1. 
C. H. K. Chen is also supported by STFC Consolidated Grant ST/X000974/1.
M.J. Owens is part-funded by Science and Technology Facilities Council (STFC) grant number ST/V000497/1 and Natural Environment Research Council (NERC) grant NE/Y001052/1. P. A. Simon acknowledges useful comments and suggestions from the space plasma group at Queen Mary University of London. We would like to thank the NASA Space Physics Data Facility and the Space Sciences Laboratory that provide access to the \emph{Wind} dataset.

\section*{Open Research}
The original spacecraft observation dataset published by NASA Space Physics Data Facility and the Space Sciences Laboratory has been accessed by the Coordinated Data Analysis Web (CDAWeb) through the Python module cdasws \cite{lin_wind_2021}.
Figures were made with Matplotlib version 3.9.1 \cite{hunter_matplotlib_2007,caswell_matplotlibmatplotlib_2021}, available under the Matplotlib license at \url{https://matplotlib.org/}.
The Python software and Jupyter notebooks used to obtain and analyse the AnEn forecasts are available on  \citeA{simon_analog_2025}.
The datasets, forecasts and analysis data are available on \citeA{simon_analog_2025-1}. 
Schematics have been built with Microsoft 365 PowerPoint.

\nolinenumbers

\end{document}